# The high-density regime of dusty plasma: Coulomb plasma

K. Avinash[1], S. J. Kalita[2], R. Ganesh[2], and P. Kaur[2]

[1]Department of Physics, Sikkim University, Gangtok, Sikkim 737102, India

[2] Institute for Plasma Research, Bhat, Gandhinagar, 382428 Gujrat, India

## Abstract

It is shown that the dust density regimes in dusty plasma are characterized by two complementary screening processes (a) the low dust density regime where the Debye screening is the dominant process and (b) the high dust density regime where the Coulomb screening is the dominant process. The Debye regime is characterized by a state where all dust particles carry an equal and constant charge. The high-density regime or the "Coulomb plasma" regime is characterized by (a) "Coulomb screening" where the dust charge depends on the spatial location and is screened by other dust particles in the vicinity by charge reduction (b) "quark" like asymptotic freedom where dust particles, which on an average carry minimal electric charge ($q_d \to 0$), are asymptotically free (c) uniform dust charge density and plasma potential (d) dust charge neutralization by a uniform background of hot ions. Thus, the Coulomb plasma is essentially a one-component plasma (OCP) *with screening* as opposed to electron plasma which is OCP without screening. Molecular dynamics (MD) simulations verify these properties. The MD simulations are performed, using a recently developed Hamiltonian formalism, to study the dynamics of Yukawa particles carrying variable electric charge. A hydrodynamic model for describing the collective properties of Coulomb plasma and its characteristic acoustic mode called the "Coulomb acoustic wave" is given.

# I. INTRODUCTION

Dusty plasma is a three-component plasma consisting of electrons, ions, and a dispersed phase of micron/sub-micron-sized particles. These particles acquire a net electric charge due to various electron and ion fluxes e.g., the thermal fluxes of plasma electrons and ions, the release of photoelectrons due to the absorption of solar UV radiation, the release of secondary electrons due to the impact of energetic primary electrons, etc. This paper considers the lab situation where particles acquire a net negative charge due to the thermal fluxes from the background plasma. This negative charge of a dust particle is shielded by the background electrons and ions. The scale length associated with this shielding, which is called the Debye shielding, is given by $\lambda_d = (T/4\pi q^2 n)^{1/2}$ where $T$ and $n$ are the plasma temperature and the plasma density. If we regard the dust particle as a spherical capacitor then the dust charge is proportional to the dust surface potential with respect to the local plasma potential $\psi_s$ i.e., $q_d = r_d \psi_s = r_d(\varphi_s - \varphi_p)$ where $\varphi_s$ is the dust potential, $\varphi_p$ is the local plasma potential, and $r_d$ is the dust size. In dusty plasma, for a given plasma density, temperature, and dust size, the dust density regime is usually specified by the normalized dust density or Havnes parameter [1,2] defined as $P = 4\pi n_d r_d \lambda_d^2$ where $n_d$ is the dust number density. In dusty plasma with a low dust density ($P \ll 1$), $\varphi_p \approx 0$. In this case, $q_d$ is maximum and is equal to the charge of an isolated dust particle $q_0$. However, in the high dust density regime ($P \gg 1$), the plasma potential $\varphi_p$ can differ significantly from zero. Hence in this regime, the average dust charge is reduced from its value $q_0$. Early experiments in dusty plasmas with micron-sized dust were conducted in the low-density regime. The high-density regime could not be accessed in these experiments probably on account of significant plasma losses on the dust surface. However, recently the high dust density regime has been accessed in dusty plasma experiments using nano-sized dust with $P$ in the range 30 to 50 [3-10]. In these experiments, a significant reduction in the value of average dust charge along with strong electron depletion in the plasma discharge was observed.

It was shown earlier in ref [11], that the dust charge reduction in the high dust density regime can be interpreted as due to the screening of the charge of a dust particle by other dust particles in the vicinity. This screening, which is called "Coulomb screening", is mediated solely by the dust particles. The background plasma plays no role in it. The scale length $\lambda_c$ associated with this screening in the high-density regime is given by $\lambda_c = (1/4\pi n_d r_d)^{1/2}$ [11]. Thus, in terms of $\lambda_c$ and $\lambda_d$, the Havnes parameter can be expressed as $P = \lambda_d^2 / \lambda_c^2$. In the low-density regime $P \ll 1$, or $\lambda_d^2 \ll \lambda_c^2$. This implies that in the low-density regime the Debye shielding will dominate over the Coulomb shielding with no charge reduction i.e., all the particles will have constant and maximum charge equal to the charge of an isolated dust particle $q_0$. On the other hand, in the high-density regime $P \gg 1$ or $\lambda_d^2 \gg \lambda_c^2$. Thus in this regime, the Debye shielding becomes ineffective. The dust particles interact with a force that is predominantly Coulombic in nature. However, this force is shielded or weakened due to particle charge reduction. This weakening of interparticle force due to charge reduction is called "Coulomb screening". The scale length of this screening is given by $\lambda_c$ and it becomes important in the high-density regime. In fact in the limit $P \to \infty$, the inter-particle force is very weak due to strong charge reduction, hence particles are asymptotically free. Thus, the dusty plasma is characterized by two complementary screening processes i.e., the Debye screening with scale length $\lambda_d$ in the low dust density limit and the Coulomb screening with scale length $\lambda_c$ in the high dust density limit. The ratio of the scale lengths is given by the Havnes parameter $P$.

The weakening of inter-particle force as particles come closer in the $P \gg 1$ regime, is reminiscent of quantum chromodynamics where quarks become asymptotically free as they come closer. Interestingly, the interrelationship of the Debye screening and the Coulomb screening in dusty plasma is somewhat similar to the interrelationship of screening in quantum electrodynamics (QED) and quantum chromodynamics (QCD). The screening in QED arises due to vacuum polarization and the force increasing as particles come closer while in QCD, due to the corresponding screening process (which is usually called the "anti-screening" in QCD parlance), the force becomes weaker as particles come closer. The situation is similar in dusty plasma. The Debye screening in dusty plasma arises due to plasma polarization and the force increases as dust particles come closer while due to Coulomb screening, the force becomes weaker as dust particles come closer at high dust densities.

In this paper, we examine the Debye and the Coulomb regimes in dusty plasmas. In particular, we focus on novel properties of the Coulomb regime. We show that the Coulomb limit of dusty plasma, or simply the "Coulomb plasma" which exits asymptotically in the limit $P \to \infty$, has following novel properties:

(1) Delocalization of the dust charge: As opposed to the Debye regime where the dust charge is constant, in the Coulomb regime the dust charge is a function of the spatial location i.e., $q_d(r)$.

(2) Dominant Coulomb screening: In the limit $P \to \infty$, the charge of a dust particle is strongly screened by other dust particles due to charge reduction.

(3) Uniform charge density and plasma potential: In the limit $P \to \infty$, the dust charge density $\rho_{dc}$ and the plasma potential $\varphi_p$ become asymptotically uniform in space.

(4) Asymptotic freedom: In the limit $P \to \infty$, $q_d \to 0$, hence in this limit, the dust particles are asymptotically free. Thus, the Coulomb plasma is microscopically almost electrically neutral ($q_d \to 0$) but is macroscopically charged $\rho_{dc} \neq 0$.

(5) Uniform ion background: In the limit of sufficiently weak Debye screening ($P \to \infty$), the hot ions form a uniform background. This uniform background of ions neutralizes the dust charge.

In the limit $P \to \infty$ thus, the electrons are almost completely depleted while the hot ions form a uniform neutralizing background. Hence, Coulomb plasma is essentially a one component plasma (OCP) with Coulomb screening and contains dust as the only dynamical species for low-frequency processes. It is different from the electron OCP [12] where electrons interact with unscreened Coulomb force. Coulomb plasma also exhibit a novel regime of collective/fluid behavior. We will show that Coulomb plasma exhibits a charecterstic acoustic mode called "Coulomb acoustic wave" where the inertia as well as the screening, both are due to the dust.

In the second part of our paper, we perform Molecular Dynamics (MD) simulations to study the Debye and the Coulomb regime of dusty plasma. In particular, we focus on the $P \to \infty$ regime and verify some of its properties listed above. The MD simulation is performed with Yukawa particles carrying variable charge i.e., using the YPVC model. The code uses a recently developed Hamiltonian formalism to study the dynamics of particles with variable charge [13]. It solves the equation of motion with the dust charging equation and the Poissons equation. The results from the code clearly show that in the limit $P \ll 1$, we recover the regime where the Debye screening dominates and all particles carry equal and constant charge equal to $q_0$. While in the limit $P \gg 1$, we obtain the Coulomb plasma regime where particles interact via screened Coulomb potential i.e., Coulomb interaction with reduced charge. The charge distribution funtion in this phase is a Gaussian around an average charge that is substantially reduced due to Coulomb screening and is in agreement with the fluid theory. The code further verifies that in limit $P \to \infty$, the average dust charge $q_d$ approaches zero but the average dust charge density as well as the plasma potential become spatially uniform. Thus the "Coulomb plasma" is a constant charge density OCP plasma with screening, which is microscopically almost neutral but macroscopically charged.

The paper is organised as follows. In Sec II, we give a unified description of the Debye and the Coulomb regimes of dusty plasma. The discussion clearly shows the two complementary screening regimes and focuses on describing new properties of the Coulomb regime. In Sec III, we give a fluid description of Coulomb plasma and study its characterstic acoustic mode. In Sec. IV we describe MD simulations of Yukawa particles with the variable charge to study collective processes of the Coulomb plasma regime characterized by substantial charge reduction, uniform charge density, and Coulomb screening. In the last section we discuss and summarise our results.

## II. THE COULOMB AND THE DEBYE SCREENING

In this section, we discuss the physics of Coulomb screening and its relation to the average dust charge reduction and the Debye screening. For this, we consider a lab situation where the dominant charging mechanism is due to the thermal fluxes of ions and electrons ($I_i, I_e$) of the background plasma. To examine the relation between Coulomb screening, Debye screening and the average dust charge reduction, we start with the dust charging equation and the quasi-neutrality condition given by

$$\frac{\partial q_d}{\partial t} = I_e + I_i \qquad (1),$$

$$\sum_\alpha q_\alpha n_\alpha = 0 \qquad (2),$$

where α denotes electrons, ions and dust. Since the dust charging time is short compared to the time scale of the dust motion, we consider the dust particles to be fully charged at all times. Under this assumption, we drop the time derivative term on the LHS of Eq. (1) in our further discussions. Also, in using the quasi-neutrality condition in Eq. (2) we have tacitly assumed that the dust charge is appropriately screened in the low as well as in the high dust density regimes by the Debye and Coulomb screenings. In the case of dust charging due to thermal fluxes of electrons and ions, the dust charge and the background plasma potential are both negative. Within the Orbit-Motion Limit (OML) approximation [1, 2], the expressions for thermal fluxes of electrons and ions are given by

$$I_e = -q\pi r_d^2 \sqrt{\left(\frac{8T_e}{\pi m_e}\right)} n_e \exp\left(-\frac{q\psi_s}{T_e}\right) \qquad (3),$$

$$I_i = q\pi r_d^2 \sqrt{\left(\frac{8T_i}{\pi m_i}\right)} n_i \left(1 + \frac{q\psi_s}{T_i}\right) \qquad (4),$$

where $\psi_s$ is the dust surface potential with respect to local plasma potential, ($q_d = r_d \psi_s$) and negative sign of the dust surface potential has been taken into account in Eq. (3) and (4). The ion and the electron densities are given by the Boltzmann relations $n_i = n_0 \exp\left(\frac{q\varphi_p}{T_i}\right)$ and $n_e = n_0 \exp\left(-\frac{q\varphi_p}{T_e}\right)$, where $n_0$ is the pristine plasma density, $\varphi_p$ is the local plasma potential and the negative sign of $\varphi_p$ has also been taken into account. Havnes *et al* [1] have shown that for given values of $n_0$, $r_d$ and the electron and the ion temperatures, Eq. (1) and (2) constitute two relations to determine average values of $\varphi_p$ and $\psi_s$ as a function of $n_d$ i.e., $\varphi_p = \varphi_p(n_d)$, $q_d = q_d(n_d)$ and with increasing dust density, $q_d$ or $\psi$ decrease while $\varphi$ increases i.e., $dq_d/dn_d \leq 0, d\varphi/dn_d \geq 0$ ($\varphi = q\varphi_p/T, \psi = q\psi_s/T$). In our paper, the numerical values of various constants will be quoted for the hydrogen plasma. Thus, in the limit of low dust density or the case of an isolated dust particle limit, P << 1, $\psi$ = 2.51 and $\varphi$ = 0. With increasing density i.e., $P \to \infty$, $\psi \to 0$ and $\varphi \to 1.88$. Dropping the time derivative and taking log on both sides of the equality, Eq. (1) can be expressed as

$$\frac{1}{2}\ln\left(\frac{m_i}{m_e}\right) = \ln(1+\psi) + \psi + 2\varphi \qquad (5),$$

where we have assumed $T_e = T_i = T$ for simplicity. We begin our discussion by considering an isolated dust particle or a low density bunch in a pristine electron-ion plasma. In this case $\varphi \approx 0$ and from Eq. (5) $\psi$ = 2.51. Next, we gradually add dust particles in the bunch. Due to increase in the density, the dust space charge in the bunch increases. As a result, the local plasma potential within the bunch becomes negative with respect to surrounding pristine plasma where $\varphi$ is still zero and a local electric field towards the bunch is creatyed. This electric field is shielded by two processes in the dusty plasma. First, it will be shielded by surrounding electrons and ions by "Debye screening". Second, due to the increase in the dust density the average dust charge in the bunch is reduced. The reduction in the bunch space charge caused by reduction in the average dust charge, reduces or shields the electric field in the bunch. This is "Coulomb screening". Hence, the perturbed quasi-neutrality condition within the bunch can be expressed as

$$q_d\Delta n_d + \Delta q_d n_d - q\Delta n_i + q\Delta n_e = 0 \qquad (6),$$

where we assumed $\Delta n_\alpha / n_\alpha \ll 1$ (α denotes dust, ion and electrons) and $\Delta q_d / q_d \ll 1$. In Eq.(6), the second term is due to the Coulomb screening while the last two terms are due to the Debye screening. Next, to derive a relation between the Coulomb screening, the Debye screening and the dust charge reduction, we express $\Delta n_d, \Delta n_i, \Delta n_e$ and $\Delta q_d$ in terms of $\Delta \varphi$ using relations $\varphi_p = \varphi_p(n_d)$, $q_d = q_d(n_d)$ and $q_d = q_d(\varphi_p)$ as follows

$$\Delta q_d n_d = -\frac{\Delta \varphi_p}{4\pi \lambda_c^2} \qquad (7),$$

$$q_d \Delta n_d = -\frac{\Delta \varphi_p}{4\pi \lambda_c^2}\left(\frac{n_d}{q_d}\frac{dq_d}{dn_d}\right)^{-1} \qquad (8),$$

$$-q\Delta n_i + q\Delta n_e = -\frac{\Delta \varphi_p}{4\pi \lambda_d^2} \qquad (9),$$

where $1/\lambda_c^2 = -4\pi(d\psi/d\varphi)r_d n_d$ and we have used $q_d = r_d \psi_s$. The derivative $d\psi/d\varphi$ may be calculated from Eq. (5) to obtain final expression of $\lambda_c$ as

$$\lambda_c^2 = \frac{1}{4\pi n_d r_d}\left(\frac{(2+\psi)}{2(1+\psi)}\right) \qquad (10).$$

Thus, with the increasing density, $\psi$ decreases from 2.51 to 0. Hence $\lambda_c$ in low-density regime is given by $\lambda_c = 0.8/\sqrt{4\pi n_d r_d}$ while in the high-density regime, it is given by $\lambda_c = (1/4\pi r_d n_d)^{1/2}$. Substituting Eq.(7)-(9) in Eq. (6) and rearranging we finally obtain

$$\left(1+\frac{n_d}{q_d}\frac{dq_d}{dn_d}\right) = \frac{\lambda_c^2}{\lambda_c^2 + \lambda_d^2} = \frac{1}{1+A} \qquad (11),$$

where $A = \lambda_d^2 / \lambda_c^2$. It should be noted that $A$ defined above is strictly not equal to Havnes parameter $P$ defined earlier. It differs from $P$ by the function of $\psi$ given in the parenthesis in Eq. (10) which is of order unity. Thus, $A$ is slightly greater than $P$ and approaches $P$ as $P \to \infty$. Hence, we will continue to use $P$ instead of $A$. The relation between the Debye screening, the Coulomb screening and the dust charge reduction is shown in Eq.(11). We first consider the low density case.

**A. Low dust density limit : Debye regime**

This regime is characterized by the limit $P \to 0$. Hence in this regime the Debye screening is the dominent screening process. The plasma consists of electrons, ions and dust. The dust charge is screened by electrons and ions. In this regime, Eq.(11) implies $dq_d / dn_d \approx 0$, thus the dust charge is independent of the dust density. There is no charge reduction. All dust particles have equal and constant charge $\psi = 2.51$ and $\varphi = 0$.

**B. High dust density limit: Coulomb regime**

This regime is characterized by the limit $P \to \infty$. Hence in this regime, the Coulomb screening is the dominant screening process. Further, from Eq. (11), we have the condition $n_d dq_d / q_d dn_d$ = -1. Several consequences follow from this condition. First, the average dust charge density $\rho_{dc} = q_d n_d$ is constant. Second, since $n_d \to \infty$ in this limit, $q_d \to 0$. Thus, the dust charge is severely reduced. In fact, the dust particles are asymptotically free in the $P \to \infty$. Third, due to the significant plasma potential in this regime $\varphi_p \to 1.88 T/q$, the electron population is almost completely depleted. Hence, the plasma consists of only two components i.e., ions and dust. Further, on account of weak Debye screening in limit $P \to \infty$, the ion density becomes uniform due to rapid thermal motion. This uniform ion population then provides the neutralizing background for negatively charged dust with uniform charge density. Thus, in this limit, the dust is microscopically almost neutral ($q_d \to 0$) but macroscopically charged ($q_d n_d \neq 0$). Also, the low frequency dynamics in this limit is only due to dust particles.

Coulomb plasma provides a new regime of collective behaviour. In the next section, we describe a fluid model of Coulomb plasma to describe some of these properties.

### III. HYDRODYNAMIC MODEL OF COULOMB PLASMA

In this section, we describe a Hydrodynamic model for collective processes in the Coulomb limit.

#### A. Hydrodynamic equations

As stated in the previous section, the low frequency dynamics in the Coulomb limit is only due to dust particles. Hence, we start with following two equations for the dust component.

$$\rho_{dm} \frac{d\vec{v}_d}{dt} = -\rho_{dc} \nabla \varphi_p \qquad (12),$$

$$\frac{\partial \rho_{dm}}{\partial t} + \nabla \cdot \left( \rho_{dm} \vec{v}_d \right) = 0 \qquad (13),$$

where $\rho_{md}$ is the dust mass density and in Eq. (12) the negative sign of the dust charge and the plasma potential has been taken into account. These two equations are supplemented with the quasineutrality condition and the dust charging equations. In the Coulomb limit ($P \to \infty$) these equations are

$$\rho_{dc} = \rho_{ci} = \text{Const.} \qquad (14),$$

$$\frac{1}{4} \frac{T}{q} \ln\left( \frac{m_i}{m_e} \right) = \psi_s + \varphi_p \qquad (15),$$

where $\rho_{ci}$ is the charge density of the uniform neutralizing background of hot ions and we have expanded the log term in Eq. (5) for small $\psi$.

Using these equations to eliminate $\varphi$ and $q_d n_d$ in Eq. (12) and (13) we obtain following equations

$$\rho_{md} \frac{d\vec{v}_d}{dt} = -\frac{\rho_{dc}^2 m_d}{r_d} \nabla(1/\rho_{md}) \qquad (16),$$

$$\frac{\partial \rho_{md}}{\partial t} + \nabla \cdot \left(\rho_{md} \vec{v}_d = 0\right) \qquad (17).$$

Thus, the Coulomb plasma is described by two variables $\rho_{md}$ and $\vec{v}_d$ which obey hydrodynamic Eq.(16) and (17) respectively. Next, using these equations we describe the characterstic acoustic mode of the Coulomb plasma.

**B. Acoustic mode**

Coulomb plasma supports a characterstic low frequency acoustic mode called the "Coulomb acoustic mode" (CAW). The characterstic acoustic mode of the Debye regime is the dust acoustic wave (14). The dispersion relation of CAW can be derived as follows. Consider a homegenous equilibrium solution of Eq.(16) and (17) having a uniform density $\rho_{md}$ and $\vec{v}_d = 0$. Linearizing Eq. (16) and (17) around this equilibrium, we obtain the dispersion relation of this mode in terms of $\rho_{md}$ as

$$\frac{\omega^2}{k^2} = C_{CAW}^2 = \frac{\rho_{dc}^2 m_d}{r_d \rho_{md}^2} \qquad (18).$$

(this mode was earlier proposed by Avinash and Shukla [11] and independently by Rao [15]. However, the names of this mode given earlier by these authors were not appropriate as these did not reflect the essential acoustic nature of the mode which arises due to imperfect shielding). A physical picture of this mode based on the imperfect coulomb shielding is discussed in the next section.

**C. A Physical picture of the Coulomb acoustic mode**

A physical picture of CAW can be given as follows. As stated earlier, in the Coulomb plasma limit, the dust charge density $(q_d n_d)$ is neutralized by a uniform background of hot ions as implied in Eq. (14). Now suppose we locally create a small bunch of dust particles by moving particles from the neighboring location. Let the number density and the average charge in the bunch be denoted by $(n_d + \Delta n_d), (q_d + \Delta q_d)$ respectively. Then the incremental space charge in the bunch is given by $q_d \Delta n_d + n_d \Delta q_d$ where the first term represents the increase in the negative space charge due to increase of negative charged particles hence is repulsive i.e., it will try to disperse the bunch. Since $\Delta q_d < 0$, the second term represents decrease in the negative space charge due to charge reduction, hence is cohesive (it is like adding ions in the bunch). In case of perfect Coulomb shielding $P = \infty$, the two terms mutually compensate each other and there is no mode. In the case of imperfect Coulomb shielding it can be easily seen from Eq. (11) that the first term which is repulsive exceeds the second term which is cohesive, by a small amount $\sim q_d \Delta n_d / P$. The electrostatic pressure due to this uncompensated space charge and concomitant electric fields along with dust inertia then drives the acoustic mode. To obtain the phase velocity of the mode we note that the electrostatic pressure due to these fields is roughly $P_E \approx q_d \psi_s n_d$. The phase velocity of the mode is then given by $C_{CAW}^2 = P_E / \rho_{md} = q_d^2 / r_d m_d$ which agrees with Eq.(18).

In the next section, we discuss molecular dynamics simulations.

# IV. MOLECULAR DYNAMICS SIMULATION OF THE DEBYE AND THE COULOMB REGIMES

Our goal in this section is to develop a quantitative model of the Debye and the Coulomb regimes of dusty plasma based on molecular dynamics simulation of Yukawa particles with variable charge. In our simulations we would specifically like to focus on the novel properties of the Coulomb regime characterized by sustantial charge reduction, the uniform charge density, and the Coulomb screening.

To address this problem we start with a situation where a dust cloud is imbedded in cubical box of volume $V$. The cloud contains $N$ dust particle of same size $r_d$. These particles carry electric charge which depends on the particle location $q_d(r)$, and are moving in the self consistent force fields. The location and the charge of $i^{th}$ particle at time $t$ is given by $\vec{r}_i$ and $q_{di}$ ($i = 1,2,3,.....N$). The MD simulation involves studying the dynamical evolutuion of this system by solving the Poissons equation, the dust charging equation and the equation of motion for $N$ particles. The surface potential of the $i^{th}$ particle $\psi_{si}(r)$ (or the charge $q_{di}$) and the plasma potential at the location of $i^{th}$ particle $\varphi_{pi}(r)$, are governed by the following equations

$$\nabla^2 \varphi_p = 4\pi \left( qn_i - qn_e - r_d \sum_{i}^{N} \psi_{si} \delta(r - r_i) \right) \quad (19),$$

$$\frac{1}{2}\ln\left(\frac{m_i}{m_e}\right) = \ln\left(1 + \frac{q\psi_{si}}{T}\right) + \frac{q\psi_{si}}{T} + 2\frac{q\varphi_{pi}}{T} \quad (20),$$

where electron and ion densities are given by the Boltzmann relations given earlier. As Poissons equation and the dust charging equation given in Eq, (19) and (20) are coupled, these are to be solved iteratively to calculate the dust surface potential and the local plasma potential for all $N$ particles at time $t$. Once all the charges and the potentials for $N$ particle are calculated at time *t*, then the new locations of these particles at time $(t+dt)$ are calculated by solving the corresponding equations of motion. A Hamiltonian formalism for studying the motion of dust particles with variable charge is presented in ref.[13]. The Hamiltonian and the equations of motion for the ensemble of *N* dust particles are given by

$$H = \sum_{i=1}^{N}\left[\frac{p_{di}^2}{2m_d} + \frac{r_d T^2}{2q^2}\left(K - \frac{(1+\psi_i)^2}{2} + \ln(1+\psi_i)\right)\right] \quad (21),$$

$$\frac{\partial p_{di}}{\partial t} = -\frac{\partial H}{\partial r_i}, \frac{\partial r_u}{\partial t} = \frac{1}{m_d}\frac{\partial H}{\partial p_{di}} \quad (22),$$

where $p_{di}, \psi_i$ are the generalized momentum. and the charge of the $i^{th}$ particle $(i=1,2,3......N)$ and *K* is a constant = 4.87 for H plasma. The derivation of Eq. (21) and (22) is given in Appendix A. These equations of motion have to be solved with the charging equations Eq. (19) and (20) to update the charge, the position, and the local plasma potential at each time step and evolve the system. The quality and the accuracy of the simulation may be checked by evaluating the ensemble average of the charge $\langle \psi_{si} \rangle$ and the plasma potential $\langle \varphi_{pi} \rangle$ from simulations and comparing it with the average charge $\psi$ and the plasma potential $\varphi$ obtained from the fluid model of charging given by Havnes et al [1]. In this model, the value of the these variables for given normalized dust density or Havnes parameter *P* are given by

$$\psi P = \sinh \varphi \quad (23),$$

$$\frac{1}{2}\ln\left(\frac{m_i}{m_e}\right) = \ln(1+\psi) + \psi + 2\varphi \quad (24).$$

Eliminating φ between these equations we have the following equation for average charge as a function of *P*.

$$\frac{1}{2}\ln\left(\frac{m_i}{m_e}\right) = \ln(1+\psi) + \psi + 2\sinh^{-1}(\psi P) \qquad (25).$$

For a good accuracy of MD simulations we should have $\langle\psi_{si}\rangle = \psi$ and $\langle\varphi_{pi}\rangle = \varphi$. The good accuracy of simulation requires that $N$ should be as large as possible; atleast of the order of $10^4$ to $10^5$ or more if possible. However, obtaining iterative solutions of Poissons equation, which is a non linear PDE, along with Eq. (20) for $N \sim 10^4$ to $10^5$ is a computationally a demanding task; even with the best Poissons solvers which are currently available. Earlier, a simple 1D solution of Eq. (19) and (20) which shows charge reduction, with point particles placed at fixed and regular intervals along a lattice for $N = 3$ and 9 has been discussed by Goertz [2]. Young *et al* [16] have performed PIC simulations of dust charging with 16 grains placed on a 2D lattice. Such calculations with small values of *N*, though instructive, are inadequate for simulation of collective processes. Clearly some simplifying approximations are necessary to solve the problem. The simplification should be such that it decouples Eq.(19) and (20). To solve this problem we propose following model for MD simulation of Yukawa particle with variable charge (YPVC)

### A. Yukawa particles with variable charge (YPVC) model

In this model the the Boltzmann relations for electrons and ions in Eq. (19) are linearized to give

$$(\nabla^2 - \lambda_d^{-2})\varphi_p = -4\pi\left(r_d\sum_i^N \psi_{si}\delta(r - r_i)\right) \qquad (26).$$

The solution of Eq. (26) is the Yukawa potential given by

$$\varphi_p = r_d\sum_i^N \psi_{si}\frac{e^{-|\vec{r}_{ij}|/\lambda_d}}{|\vec{r}_{ij}|} \qquad (27).$$

where $|\vec{r}_{ij}| = |\vec{r}_i - \vec{r}_j|$. Usually, the solutions of Poisson equation in Eq. (26) are obtained for Yukawa particles carrying a constant charge. Here, in Eq. (27), we have a linear solution of Possion equation for Yukawa particles carrying a variable charge which depends on the local position. We will construct a MD simulation model based on this solution of Poissons equation given in Eq. (27).

The wisdom of linearization in Eq. (26) may be questioned due to the reason that it is only valid for $q\varphi_p/T < 1$ which is valid in the low-density Debye regime while we are interested in the simulation of the high density Coulomb regime where the approximation $q\varphi_p/T < 1$ is invalid. In the high density regime $q\varphi_p/T \geq 1$. We will return to this question shortly.

Eliminating $\varphi_{pi}$ in Eq. (20) by carrying out the summation over all particles except the $i^{th}$ particle, in Eq. (27), we obtain following set of $N$ equations

$$\frac{1}{2}\ln\left(\frac{m_i}{m_e}\right) = \ln(1+\psi_i) + \psi_i + 2r_d \sum_{j \neq i}^{N} \psi_j \frac{e^{-|\vec{r}_{ij}|/\lambda_d}}{|\vec{r}_{ij}|} \quad , (i=1,2,3,.....N) \qquad (28),$$

where $q\psi_{si}/T = \psi_i$. For given locations $\vec{r}_i\ (i=1,2,3,.....N)$ of $N$ dust particles at time $t$, the mass ratio and the particle size $r_d$, the set of Eqs. (28) constitute a set of $N$ non-linear coupled algebraic equations which can be solved to obtain values of $N$ dust charges $\psi_i\ (i=1,2,3,.....N)$ at time $t$. Substitutiting the values of dust charges $\psi_{si}$ in Eq. (27), the corresponding plasma potential of the charge configuration $\varphi_{pi}\ (i=1,2,3,.....N)$ can also be calculated. Substituting values of $\psi_{si}$ and $\varphi_{pi}$ in equations of motion Eq. (21) and (22), new positions $(\vec{r}_i + d\vec{r}_i)$ at time $(t+dt)$ of particles can be calculated and the temporal evolution of the system can be studied.

Next, we address the issue of the suitability of this model which consists of solving Eq. (28), along with (21) and (22) for the simulation of the high density Coulomb regime. A simple way to acertain this could be to take the fluid limit of this model and check whether it shows charge reduction and Coulomb screening in the limit of large $P$. The fluid limit of YPVC model can be taken by replacing $\psi_{si} \rightarrow \psi$ and $\sum_{i}^{N} \rightarrow n_d \int dV$ and performing the space integration in Eq. (27) and (28) to give $\varphi = \psi P$ and

$$\frac{1}{2} \ln\left(\frac{m_i}{m_e}\right) = \ln(1+\psi) + (1+2P)\psi \qquad (29).$$

Comparing the last term in Eq. (29) with the corresponding last term Eq. (25) we see, as expected, that the fluid limit of the present model corresponds to using linearized quasi neutrality condition in the Havnes model given in Eq.(23) and (24). Thus the two models are expected agree with each other in the Debye limit $P \rightarrow 0$ i.e., $\psi \rightarrow 2.5$ and $\varphi \rightarrow 0$ in both models. However, interestingly, the structure of Eq. (29) is such that it agrees with Havnes model even in the $P \rightarrow \infty$ limit! Thus, Eq. (29) also shows charge reduction and Coulomb screening. In the limit $P \rightarrow \infty$, Eq. (29) gives $\psi \rightarrow 0$ while $\varphi \rightarrow 1.88$ exactly as in Havnes model. Thus the YPVC model in Eq. (29) approaches the Havnes model in Eq. (25) in $P \rightarrow 0$ as well as $P \rightarrow \infty$ limits  Hence YPVC model should be adequate for the  simulation of Debye regime (as expected) and more importantly of Coulomb regime as well. It should be noted that the two models will differ with each other for interneidate values of $P$ of order unity. In this range the YPVC model may only be qualitatively correct. But this range of $P$ is not of interest in this paper.

In the Coulomb limit, Eq. (28) can be further simplified into equations with some nice features. In this limit $\psi \ll 1$, hence the  log term in Eq. (28) can be linearized to give

$$\frac{1}{4}\ln\left(\frac{m_i}{m_e}\right) = \psi_i + r_d \sum_{j \neq i}^{N} \psi_j \frac{e^{-|\vec{r}_{ij}|/\lambda_d}}{|\vec{r}_{ij}|} \quad , \quad (i = 1,2,3,.....N) \tag{30}$$

Eq. (30) constitutes a set of $N$ linear equations for $N$ charges $\psi_i, (i = 1,2,3,.....N)$. For given location $\vec{r}_i$ of these charges, values of $\lambda_d$ and $r_d$, and the mass ratio, these equations can be solved to obtain values of $\psi_i$ at time $t$. These equations can be expressed in matrix form as

$$AX = B \tag{31}$$

where $X$ is the column matrix of $N$ unknown elements $\psi_i$ $(i = 1,2,3,.....N)$, $B$ is another column matrix of $N$ elements which are all $= (1/4)\ln(m_i/m_e)$ and $A$ is a square, symmteric matrix of $N$x$N$ elements $b_{ij} = (r_d/|\vec{r}_{ij}|)\exp(-|\vec{r}_{ij}|/\lambda_d) = b_{ji}$ $(i \neq j)$, with all $N$ diagonal elements $b_{ii}$ equal to unity. The matrix $A$ can be inverted to give the solution of Eq. (31) as $X = A^{-1}B$ which then gives the values of $N$ dust charges $\psi_i$ at time $t$. As can be appreciated, solving $N$-coupled nonlinear algebraic equations or $N$-coupled linear algebraic equations of the present model is computationally much simpler than solving a set of N-coupled nonlinear partial differential equations of the fully nonlinear model in Eq. (19) and (20). The YPVC model thus provides a computationally simple and a feasible method of simulating especially the Coulomb limit of dusty plasma. Thus in our model we solve Eq. (28) or (30) along with equations of motion in Eq. (21) and (22) to study the dynamical evolution of the system. In the next section, we construct a molecular dynamics simulation code based on the YPVC model. The results of this simulation will be compared with the corresponding fluid theory in Eq.(29).

Next, we discuss the normalization scheme used in our simulations. The simulation is performed in a cubical box of dimension $L$ and volume $V$. The box containes $N$ dust particles at given locations i.e., the $i^{th}$ dust particle is located at $\vec{r} = \vec{r}_i$ ($i = 1,2,3......N$) at time $t$. The $i^{th}$ particle carries a normalized electric charge $\psi_i$. The velocity of the $i^{th}$ particle at time, is given by $\vec{v} = \vec{v}_i$ In our simulations, we solve the charging Eq. (28) or (30) along with equations of motion of N dust particles. These equations are given by

$$m_d \frac{d^2 \vec{r}_i}{dt^2} = -r_d \psi_{si} \frac{d}{d\vec{r}_i} r_d \sum_{j \neq i}^{N} \psi_{sj} \frac{\exp(-|\vec{r}_{ij}|/\lambda_d)}{|\vec{r}_{ij}|} , \quad (i = 1,2,3......N) \quad (32).$$

In these equations, we have eliminated $\varphi_{pi}$ from Eq. (27). The total energy of the system $E$, which is the sum of the potential and the kinetic energy, is a constant of motion given by the Hamiltonian $H$ given in Eq. (21). In our simulations we use following normalization scheme. All lengths will be normalized by $a$ which is the mean particles distance given by $a = (3/4\pi n_d)^{1/3}$ where $n_d$ is the global particle density given by $n_d = N/V$ and time is normalized by $\tau = (m_d a q^2 / T^2)^{1/2}$ which is the typical time scale of particle motion. In terms of these units we have $\bar{n}_d = 3/4\pi$, $\bar{\lambda}_c = \lambda_c / a, \bar{\lambda}_d = \lambda_d / a, \kappa = 1/\bar{\lambda}_d$. $|\vec{r}_{ij}| = |\vec{r}_i - \vec{r}_j|/a$, $\bar{r}_d = r_d / a = 1/3\bar{\lambda}_c^2$ and the Hamiltonian is normalized as $\bar{H} = H/(aT^2/q^2)$. The normalized equations to be solved in simulations are

$$\frac{1}{4}\ln\left(\frac{m_i}{m_e}\right) = \frac{1}{2}[\ln(1+\psi_i)+\psi_i]+\frac{1}{3\bar{\lambda}_c^2}\sum_{j \neq i}^{N} \psi_j \frac{e^{-\kappa|\vec{r}_{ij}|}}{|\vec{r}_{ij}|} \quad (i = 1,2,3,......N) \quad (33).$$

In the Debye regime these equations are non linear in $\psi_i$ while in the the Coulomb regime where $\psi_i \ll 1$ these equations reduce to a set of linear equations given by

$$\frac{1}{4}\ln\left(\frac{m_i}{m_e}\right) = \psi_i + \frac{1}{3\overline{\lambda}_c^2}\sum_{j\neq i}^{N}\psi_j \frac{e^{-\kappa\left|\vec{r}_{ij}\right|}}{\left|\vec{r}_{ij}\right|}, \quad (i=1,2,3,.....N) \quad (34).$$

The normalized equations of motion are given by

$$\frac{d^2\vec{r}_i}{dt^2} = -\frac{\psi_i}{9\overline{\lambda}_c^4}\frac{\partial}{\partial \vec{r}_i}\sum_{j\neq i}^{N}\psi_j \frac{e^{-\kappa\left|\vec{r}_{ij}\right|}}{\left|\vec{r}_{ij}\right|} \quad (i=1,2,3,.....N) \quad (35).$$

The total energy of $N$ particles in normalized units is given by

$$\overline{E} = \sum_{i=1}^{N}\left[\frac{v_i^2}{2} + \frac{1}{6\overline{\lambda}_c^2}\left(K - \frac{(1+\psi_i)^2}{2} + \ln(1+\psi_i)\right)\right] \quad (36),$$

where $\vec{v}_i = d\vec{r}/d\bar{t}$. These equations are solved for given value of $P$ which is determined by choosing appropriate values of $\overline{\lambda}_c, \overline{\lambda}_d$, the mass ratio $(m_i/m_e)$ and given values of the positions $\vec{r}_i$ and the velocities $\vec{v}_i$ $(i=1,2,3,.....N)$ of $N$ particles at the starting time $t$. The Debye and the Coulomb regimes are be accessed by varying $P$. The ensemble averages which can be obtained from simulations are those of charge $\langle\psi\rangle$, the potential $\langle\varphi\rangle$, and local dust charge density $\langle\rho_{dc}\rangle$ etc. Results of our simulations will show that, consistent with fluid theory, the values of these ensemble averages depend only on value of $P$ and not on individual values of $\overline{\lambda}_c$ and $\overline{\lambda}_d$. The values of these averages can be compared with the fluid theory results obtained from Eq. (29) to acertain the accuracy of the simulations.

**B. MD simulation using the YPVC model**

We first describe our simulation box. A schematic diagram of the cubical simulation box of side $L$ is shown in Fig.1. Consistent with the normalization that lengths are normalized with average inter-particle distance *a*, the particles are distributed within this simulation box with a global density $3/4\pi$. All the three directions are taken to be periodic. We have performed our simulations in boxes of two sizes, $L = 32.27$ and $69.38$. The reason for taking simulation boxes of two sizes is to avoid doing the Ewald sum which involves summing over interactions with image configurations resulting from periodic boundary conditions. In the Debye regime, where $\overline{\lambda_d} < \overline{\lambda_c}$, the range of interaction among the grains is relatively smaller. Therefore a very large system size is not required, even for a triply periodic case. On the other hand, in the Coulomb regime where $\overline{\lambda_c} < \overline{\lambda_d}$, the range of interaction is larger hence a comparatively larger box is required to make interaction with image configurations weaker and avoid doing Ewald sum. Consistent with the global density of $3/4\pi$, the no of particles in the two boxes are 8000 and 79507 respevtively. A Gaussian thermostat is used to thermalize the dust grains in the beginning of the simulation. The normalised inverse temperature $\Gamma$ of the ensemble of particles is given by the inverse of the normalized temperature $\Gamma = 1/\overline{T_d}$. The parameter $\Gamma$ defined here should be carefully distinguished from the $\Gamma$ of the usual constant charge simulations. In both cases, $\Gamma$ is the inverse temperature but in the present case as the particle charge is not constant hence the normalization is different. In the simulation box, we solve Eq. (33) or (34) at time *t* for charges $\psi_i$ of *N* particles.

There are a number of methods available to solve these equations. In our simulations we solve matrix equation Eq. (33) by Conjugate gradient method. The full nonlinear Eq. (33) is solved in two steps. In the first step, the *N* nonlinear equations are solved to calculate *N* terms of the form $(\ln(1+\psi_i)+\psi_i)$, $i = 1,2,3.........N$. In the second step, Newton Raphson method is used to obtain $\psi_i$ from these nonlinear terms. We have used the upgraded version of the MPMD-3D code to solve these equations along with the Equation of motion Eq. (35) for various initial sconditions. This code is available in CPU as well as GPU versions.

## III. SIMULATIONS RESULTS

In our simulations, we study the Debye and the Coulomb regimes of the dusty plasma. The Debye regime where particles carry constant charge hitherto has been well studied. Hence in our simulations we have focussed on novel properties of the Coulomb regime. We first present results of the Debye regime simulations.

### A. Debye regime

To test our simulation code in this regime, we first check energy conservation given in Eq. (36). For this we perform a simulation run with following parameters $N = 8000$, $\overline{\lambda}_d = 1$, $\overline{\lambda}_c = 10$, $P = 0.01$ and $\Gamma = 10$. The run is initiated with particles arranged in a lattice along three directions. As we evolve the system microcanonically, the system remains in equilibrium. In Fig.2, we present the total energy per particle $\overline{E}/N$ calculated from Eq.(36). We observe that the system remains in equilibrium and the total energy (per particle) remains constant. In the same figure the kinetic and the potential energies per particle are also plotted. The plots show that, as expected, in the Debye regime the potential energy is very small $\approx 10^{-4}$ as compared with the kinetic energy per particle. Therefore the total energy is close to the kinetic energy per particle which is $\approx 1/\Gamma = 0.1$ as shown in Fig.2. Since the system is in equilibrium, we can choose any data points as point of observation beyond, $t = \tau$. Most of the diagnostics performed in this paper are at time $t = 2.90\tau$.

Next, we make simulaton run for two cases, viz, ($\overline{\lambda}_d$, $\overline{\lambda}_c$, $P$) = (1,5, 0.04) and (1,10, 0.01). The inverse temperature $\Gamma = 10$ for both cases. As stated before, in the Debye regime, we are able to perform successful runs with only 8000 particles. From these runs, we calculate the average charge and compare with the values obtained from the fluid approximation given in Eq (29). Clearly, now since our code allows delocalization of charge [13], particles at different locations at time *t* will have different values of charges. The mean value of charge from simulations are $\langle \psi \rangle$ = 2.3741 and 2.4688 for (1,5, 0.04) and (1,10, 0.01) respectively. The average values of charge for the two values of *P* from Eq. (29) are $\psi$ = 2.3594 and 2.4672 respectively. Thus the simulation values are in good agreement with fluid approximation values with under 2% error. As expected, for smaller values of *P* the values of average charge is closer to the value of the charge of an isolated particle = 2.51. In Fig. 3 the temporal fluctuation of charge around the mean value for the two cases is shown. Next, we consider the Coulomb regime.

**B. Coulomb regime**

As in the case of the Debye regime, we first check the consistency of the code by verifying the energy conservation in the Coulomb regime. To access the Coulomb regime we perform simulation runs with the following parameters: *N* = 79507, Γ = 120, $\overline{\lambda_c}$ = 1, 2, 3, $\overline{\lambda_d}$ = 10 and *P* = 100, 25, and 11.11. The mean charge obtained from the simulation are $\langle \psi \rangle$ = 0.0204, 0.0791, 0.1695 respectively. In Fig.4 we plot the average charge as a function of time *t*. These plots show that the system is in equilibrium as mean values of mean charges are time-independent. The total energy, kinetic energy, and potential energy per particle for $\overline{\lambda_c}$ = 3, $\overline{\lambda_d}$ = 10 as a function of time are shown in Fig.5. The total energy per particle is conserved up to four places of decimal. The consistency of simulation can be seen from the fact that the potential energy per particle in the Coulomb regime, which earlier was negligible in the Debye regime, has now become significant ≈ 0.08. To further check the energy conservation, we calculate the total energy per particle for the case $\overline{\lambda_c}$ = 3, $\overline{\lambda_d}$ = 10 by noting that since Γ = 120 hence the kinetic energy per particle will be ≈ 1/Γ = 0.008. The potential energy per particle can be calculated from the A-potential given in Eq. (A8) of the Appendix A. Thus substituting ψ = 0.17 in Eq. (A8) gives potential per particle ~ 0.08. This value agrees fairly well with the simulation value of 0.08. The calculated total energy per particle $\overline{E}/N$ then is ~0.088 which is reasonably close to the simulation value of 0.09. The charge distribution function $f_c(\psi)$ for three cases obtained from the simulation and the Gaussian fit to $f_c$ are shown in Fig.6. The figure shows that the Gaussian fit to $f_c$ is very good implying thereby that in the Coulomb regime, $f_c$ is well approximated by a Gaussian distribution function. The fact that in the Coulomb regime $f_c$ is close to a Gaussian could be understood possibly from the Central limit theorem [17] which states that values of variables which are determined by a large number of random processes are Gaussian distributed. In the Coulomb regime, the charge of a particle is a result of a large number of random process i.e., the random locations of all the other particles. Hence, the charge of a particle which is the result of these random processes is expected to be Gaussian distributed. In Fig. 7 we have plotted $f_c$ for different values of Γ but same $P$ to show that the mean charge is independent of Γ. The figure clearly shows that with decreasing Γ the distribution function becomes broader as expected but the mean charge, which depends only on $P$, remains constant.

Next, we study the transition to the Coulomb regime by increasing $\overline{\lambda_d}$ while keeping $\overline{\lambda_c}$ constant. We make simulations runs for $\overline{\lambda_c} = 1$ and $\overline{\lambda_d} = 3.3, 5, 10$. Note that in these runs though values of $\overline{\lambda_c}$ and $\overline{\lambda_d}$ are different, the value of $P$ is the same as in Fig.4 namely P = 10, 25, and 100. The inverse temperature in all three cases is 120 and $N = 79507$. The values of the mean charge in three cases obtained from simulations are $\langle\psi\rangle = 0.1690, 0.0740$, and $0.0204$ respectively while the mean charge from Eq. (29) are $\psi$ = 0.1715, 0.0724, and 0.0186 respectively. Thus the agreement between simulation and theory is good with an error within 10 %. These results clearly show that with increasing $P$ the Coulomb screening becomes important causing significant charge reduction. Out of the three cases studied, the last case with $P$ = 100 and strongly reduced charge ~ 0.02 is well within the Coulomb regime. Next, we calculate the average potential from simulation. For P = 10, 25, and 100, $\langle\varphi\rangle$ = 1.71, 1.806, and 1.8695 respectively. While from from Eq. (29) these values are $\varphi$ = 1.715, 1.81, and 1.86 respectively. Thus for potential the agreement between sumulation and theory is very good with error less than 1%. Also, with increasing P, we see the trend of $\langle\varphi\rangle \to$ 1.88 which is expected from Eq.(29).

Another important feature of the Coulomb plasma is the uniform dust charge density. From Eq. (29), the dust charge density in the normalized form is given by $\psi P$. In order to obtain this quantity from simulations, we calculate the charge density $\rho_{dc} = Q_d(r)/v = \psi \bar{n}_d$ where $Q_d(r)$ is the total dust charge in a small cell around any point $\vec{r}$ in the computational box and $v$ is the volume of the cell ($\psi(r)$ is the local average charge and $\bar{n}_d(r)$ is the local dust density). This quantity calculated from the simulation is multiplied by the constant $4\pi \bar{r}_d \bar{\lambda}_d^2$ to obtain the normalized charge density $\psi P$. As discussed earlier, from Eq. (29), $\psi P = \varphi$ and $\varphi \to 1.88$ as $P \to \infty$. Hence in the limit $P \to \infty$, the normalized charge density becomes spatially uniform everywhere and is equal to 1.88. We see evidence of this in our simulation

In Fig.8 we show plots of $\rho_{dc}(x,z)$ in $y = 0$ plane for $\bar{\lambda}_c$ = 1,2, and 3, and $\bar{\lambda}_d$ = 10 and P =100, 25 and 11.11. The plots clearly show that fluctuations in the charge density decrease as we go we go into the Coulomb regime by increasing $P$.

In Fig.9 we show "charge density cubes". These are full 3D plots of the local charge density $\rho_{dc}(x,y,z)$ in the simulation box. The value of charge density is color-coded. The three charge density cubes are for $\bar{\lambda}_c = 1, \bar{\lambda}_d = 3.3, 5, 10$ and $P = 10, 25$ and $100$ respectively. The cubes for $P = 10$ shows non uniform charge density with large fluctuations around the mean as indicated by the presence of different colors in the cube. With increasing $P$ the fluctuations in the local charge density reduce. In the cube with $P = 100$ the charge density is very nearly uniform as indicated by the presence of only red color in Fig. 9. From simulations the mean local charge density for $P = 10, 25$ and $100$ is $0.0384, 0.0172$ and $0.0044$ respectively. Multiplying these by $4\pi \bar{r}_d \bar{\lambda}_d^2$ we get $\psi P = 1.61, 1.80$, and $1.84$ which again shows that the normalized charge density becomes spatially uniform and approaches its asymptotic value of $1.88$ everywhere in the Coulomb limit.

Since $\bar{\lambda}_d$ is large, the Debye screening is ineffective in the state with $P = 100$. Particles are in this state are interacting with Coulomb interaction which is very weak due to strong Coulomb screening and consequent charge reduction. And the charge density is nearly spatially uniform $\approx 1.84$. Thus the state with $P = 100$ may be quite close to the regime of "Coulomb plasma" mentioned earlier. It is interesting to note that if we were working with full model in Eq. (19) and (20), then the asymptotic value of the normalised charge density in the Coulomb plasma limit would have been $\psi P = \sinh\varphi = 3.2$

In the end we would like to emphasize that even though the Havnes parameter $P$ does not appear directly in our governing equation Eq.(34), the ensemble averages of our simulations e.g., average values of dust charge, plasma potential, dust charge, as expected from the fluid theory, depend only on the value of the Havnes parameter $P$ and not on the individual values of $\lambda_c$ and $\lambda_d$. To support this, we point out that from our simulations, we have $\langle \psi \rangle = 0.0791$ for $\bar{\lambda}_c = 2$, $\bar{\lambda}_d = 10$, $P = 25$ and $\langle \psi \rangle = 0.0740$ for $\bar{\lambda}_c = 1$, $\bar{\lambda}_d = 5$, $P = 25$. As expected, since $P = 25$ in both cases, the value of the average charge are equal (within an error of ~5%), though the values of screening scale lengths in the two cases are different.

## IV. SUMMARY AND DISCUSSIONS

To summerise, we have shown that depending on the screening process, the dusty plasma exits in two distinct regimes i.e., the Debye regime with scale length $\lambda_d$ and the Coulomb regime with scale length $\lambda_c$. The Debye regime is the low dust density regime ($P = \lambda_d^2/\lambda_c^2 < 1$) where the dust charge is screened by the Debye shielding (due to background plasma) and all particle carry equal and maximum electric charge $q_0$ for given dust size and plasma parameters. The Coulomb regime, on the other hand, is the high dust density regime ($P = \lambda_d^2/\lambda_c^2 > 1$) where the dominent screening mechanism is the Coulomb ecreening. In this screening process, the dust charge is shielded by surrounding dust charges by charge reduction. In this process dust particles have different charges depending on their spatial location. This 'delocalization' of the dust charge implies that it is no longer an attribute of a dust particle like its mass [13]. In Coulomb screening, the background plasma plays no role. In our discussions we have focussed on delineating and discussing novel properties of the Coulomb regime of dusty plasma in the limit $P \to \infty$. Specifically, using the quasi neutrality condition and the dust charging equation, we have shown that the Coulomb plasma is characterized by uniform dust charge density $\rho_{dc}$ and "quark" like asymptotic freedom where $q_d \to 0$ and dust particles are asymptotically free. In this limit, the electrons are depleted and ions form a hot uniform background which neutralizes the dust charge. Hence the "Coulomb plasma" is a one-component (dust) plasma *with screening*. It is pertinent to note that the usual one-component plasma consists of electrons without screening. Further, a Hydrodynamic description of Coulomb plasma is given which shows the existence of a characteristic acoustic mode called "Coulomb acoustic mode" A physical picture of this mode shows that this mode arises due to imperfect Coulomb screening analogous to Dust acoustic wave in the Debye regime which arises due to imperfect Debye screening.

In the second part of our paper, we have performed molecular dynamics simulations foucssing on Coulomb plasma and verified some of its properties. It is argued that the complete problem for this purpose requires iterative solutions of Poissons eqation and charging equation for a large number ($\sim 10^5$ to $10^6$) of particles which is a computationally demanding task even with the best Possions solver that are currently available.Hence a simpler and computationally tractable model based on Yukawa particles with variable charge (YPVC) is used We have shown that this model tends to Havnes charging model in $P \to 0$ and $\infty$ limits. Hence it is appropriate for simulaton of Coulomb plasma in the $P \to \infty$ limit. For integrating the equations of motion the code uses a recently proposed Hamiltonian formalism for dynamics of particles with variable charge [13]. The results shows Coulomb screening and dust charge reduction for values of $P > 1$. Particles have different charge depending on the spatial location. The charge distribution function is approximated by a Gaussian to high degree of accuracy. The results for average values of the dust charge, the dust charge density and the plasma potentail agree well with corresponding fluid theory given in Eq. (29). Results of simulations with increasing values of $P$ show a tendency of homogenization of the dust charge density. For $P = 100$, the Coulomb screening and the consequent dust charge reduction are significant, hence dust particles are asymptotically free, and the dust charge density is very nearly homogeneous with mean value ~1.84 which is close to the asymptotic value 1.88. Thus the dusty plasma in this state may be a good approximation to the Coulomb plasma. For larger values of $P > 100$, our code is beset with convergence problems. One of the probable solutions to this problem is parallelization of the code which is in progress and the results for higher values of $P$ from this upgraded code will be reported in future.

As stated earlier one of the possible ways to achieve the high dust density regime in lab experiments is via nano-dusty plasma experiments. Several such experiments have obtained values of $P$ in the range of a few tens and a concomittant significant reduction in the average dust charge [3-10]. Some of these experiments have reported observations of wave activity which was attributed to the presence of dust density waves. Since these experiments are in the high dust density regime with $P$ close to a few tens, these waves could very well be the Coulomb acoustic waves discussed in this paper.

In our analysis we have made a number of assumptions i.e., Boltzmann distribution for electrons as well as ions and equal electron and ion temperatures. The assumption of thermal equilibrium for the background plasma is suitable for astrophysical situations [1, 2] especially within Giant molecular clouds. Within these clouds gas and dust is distributed within cores and clumps which are local regions of high density of gas and dust and are believed to be the locations of star/planet formation [18,19]. The Hydrogen gas in the cloud is ionised especially in the HII regions where temperatures could be in excess of a eV. The dust in this plasma background gets electrically charged. The charged dust then gravitationally collapses and accumulates in clumps to form compact dust cores which could be in the Coulomb regime with $P \sim$ a few tens. In lab experiments, the assumption of Boltzmann relation for electrons and ions with equal temperatures is also valid in Q machines. In dusty plasma Q machine experiment in Iowa [20], the Coulomb regime was achieved with $P \sim 1\text{-}5$ and significant dust charge reduction. In other lab experiments in dusty plasma e.g., DC glow discharge, RF discharge, the electron temperature usually is greater than the ion temperature. Our analysis can be easily modified to take into account this situation. However, the assumpmtion of Boltzmann distribution for ions, especially in nano dusty plasma experiments where large ion flows are observed, is questionable. Melzer and Goree [21] have proposed a constant ion density model to argue that ion flow into dust cloud is better described by flux conservation. Thus for $\varphi \sim 1$, $P \sim 1$, the ion density within the cloud is constant and equal to the density away from the cloud i.e, $n_i \approx n_0$. Havnes et al [22] have also argued that in case of dusty plasma with negative plasma potential, the ions pass through dust cloud maintaining a constant density. The flux of these constant density ions then could provide neutralizing background for dust charge in the $P \rightarrow \infty$ in the limit. However, these issues require further considerations and discussions.

As we are using periodic boundary conditions, the accuracy of the MD simulations depend on how we deal with the periodic boundaries. In the ``Debye'' regime, the range of interaction among the dust grains is small. Therefore, even with a smaller system size, we can simulate the "Debye" regime correctly using periodic boundary conditions. However, in a "Coulomb" plasma, the range of interaction potential is long. The periodic boundary condition creates arrays of replicated systems to mimic an infinite system. Therefore, the long-range nature of the interaction potential takes contribution from the nearby replicas. To avoid this one needs to consider a very large system size or take into account the "Ewald sum". In the present work, we have opted for the first option and chosen a larger system size to study Coulomb plasma.

**APPENDIX A: HAMILTONIAN FOR PARTICLES WITH VARIABLE CHARGE**

The equation of motion of a dust particle is given by

$$m_d \frac{d^2 \vec{r}}{dt^2} = -q_d \nabla \varphi_p \qquad (A1).$$

To obtain potential energy $U$ from this equation we substitute $q_d = r_d \psi$, $\varphi = q\varphi_p/T$, $\psi = q\psi_s/T$. Wth these substitutions Eq. (1) gives

$$m_d \frac{d^2 \vec{r}}{dt^2} = -\left(\frac{r_d T^2}{q^2}\right)\psi \nabla \varphi \qquad (A2).$$

In Eq. (A2), $\psi$ is a function of $\varphi$ through the charging equation given in Eq. (5) in the text. Hence, the potential energy of the dust particle is given by

$$U = \frac{r_d T^2}{q^2} \int \psi d\varphi \qquad (A3).$$

Using Eq. (5) given in the text, to perform the integration in Eq. (A3) we obtain following expression for $U$

$$U = \frac{r_d T^2}{2q^3}\left[ K - \frac{1}{2}(1-\psi)^2 + \ln(1-\psi) \right] \qquad (A4),$$

where $K$ is the constant of integration. It is determined from the condition that $U = 0$ when the plasma potential $\varphi = 0$. In this case, $K$ is given by the following equation

$$K = -\ln(1-\psi_0) + \frac{1}{2}(1-\psi_0)^2 \qquad (A5)$$

where $\psi_0$ is the solution of the following equation

$$\left(\frac{m_i}{m_e}\right)^{1/2} = (1-\psi_0)\exp(-\psi_0) \qquad (A6).$$

For H plasma $\psi_0 = 2.51$ and $K = 4.87$. The kinetic energy of the particle is given by $m_d v_d^2/2 = p_d^2/2m_d$ where $p_d$ is the generalized momentum. Hence the Hamiltonian $H$ of the particle is given by

$$H = \frac{p_d^2}{2m_d} + \frac{r_d T^2}{2q^2}\left[ K - \frac{1}{2}(1-\psi)^2 + \ln(1-\psi) \right] = q_0 \Phi_A \qquad (A7),$$

where the dust normalized charge $\psi$ is a function of local position i.e., $\psi = \psi(r)$ and $\Phi_A$ is the A potential given by

$$\Phi_A = \frac{T}{2q\,\psi_0}\left[K - \frac{1}{2}(1-\psi)^2 + \ln(1-\psi)\right] \qquad (A8).$$

Generalizing it a collection of $N$ particles

$$H = \sum_{i=1}^{N}\left[\frac{p_{di}^2}{2m_d} + \frac{r_d T^2}{2q^2}\left(K - \frac{(1+\psi_i)^2}{2} + \ln(1+\psi_i)\right)\right] \qquad (A10),$$

where the position and momentum of $i^{th}$ particle are given by $\vec{r}_i$ and $\vec{p}_{di}$ and $\psi_i = \psi_i(\vec{r}_i)$ $(i = 1,2,3,.....N)$. The corresponding equations of motion are given by

$$\frac{\partial p_{di}}{\partial t} = -\frac{\partial H}{\partial r_i},\; \frac{\partial r_u}{\partial t} = \frac{1}{m_d}\frac{\partial H}{\partial p_{di}} \qquad (A11)$$

\*\*\*\*\*\*\*\*\*\*\*\*\*\*\*\*\*\*\*\*\*\*\*\*\*\*\*\*\*\*\*\*\*\*\*\*\*\*\*\*\*\*\*\*\*\*\*\*\*\*\*\*\*\*\*\*\*\*\*\*\*\*\*\*\*\*\*\*\*\*\*\*\*

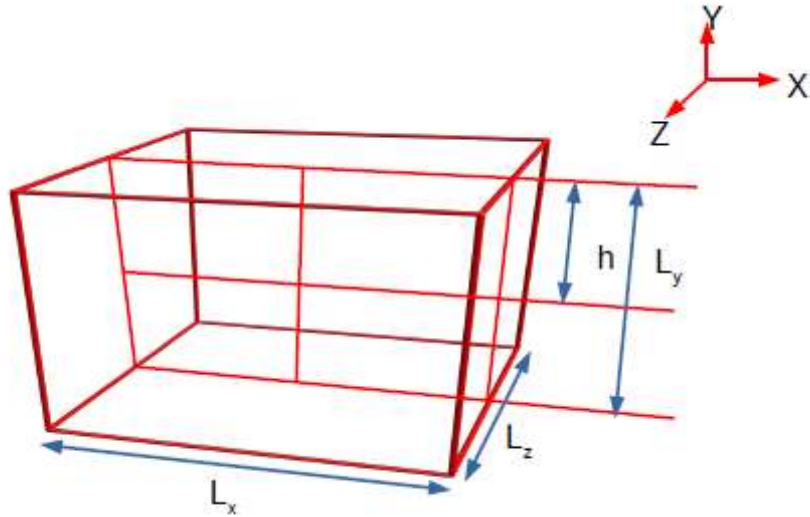

**Fig.1.** Schematic diagram of the simulation box of dimensions $(L_x, L_y, L_z)$ and volume *V*. The particles are distributed in the box with a global density of $3/4\pi$. All the three directions are taken to be periodic. For Debye regime simulations $V = (32.27)^3$ and for the Coulomb regime $V = (69.38)^3$.

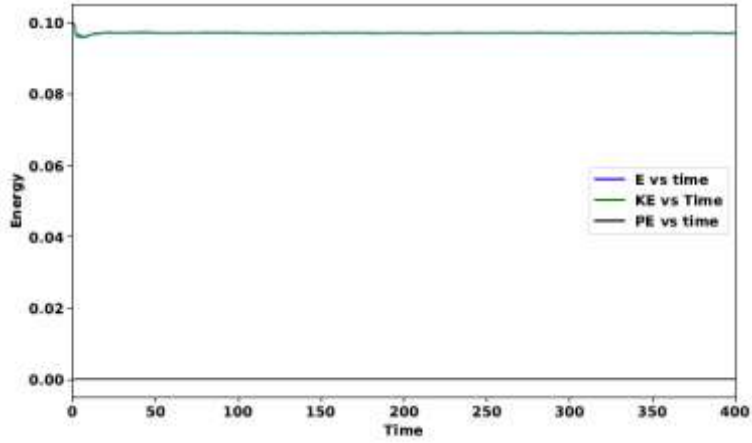

**Fig.2**. The diagram shows the total energy, the kinetic energy and the potential energy per particle for the Debye regime $P = 0.01$ ($\overline{\lambda}_d = 1$ and $\overline{\lambda}_c = 10$) and $\Gamma = 10$. As expected in this regime the potential energy is small as compared to kinetic energy which is equal to $\approx 1/\Gamma = 0.1$. The energy conservation is achieved up to 3$^{rd}$ place after 10 to 15 $\tau$.

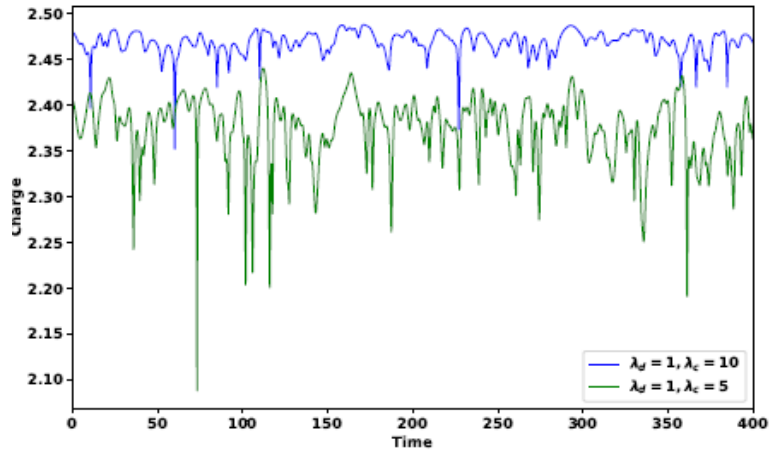

**Fig.3.** The temporal fluctuation of average charge for $P$ = 0.01 and 0.04. The fluctuations are around the mean values of 2.37 and 2.47,

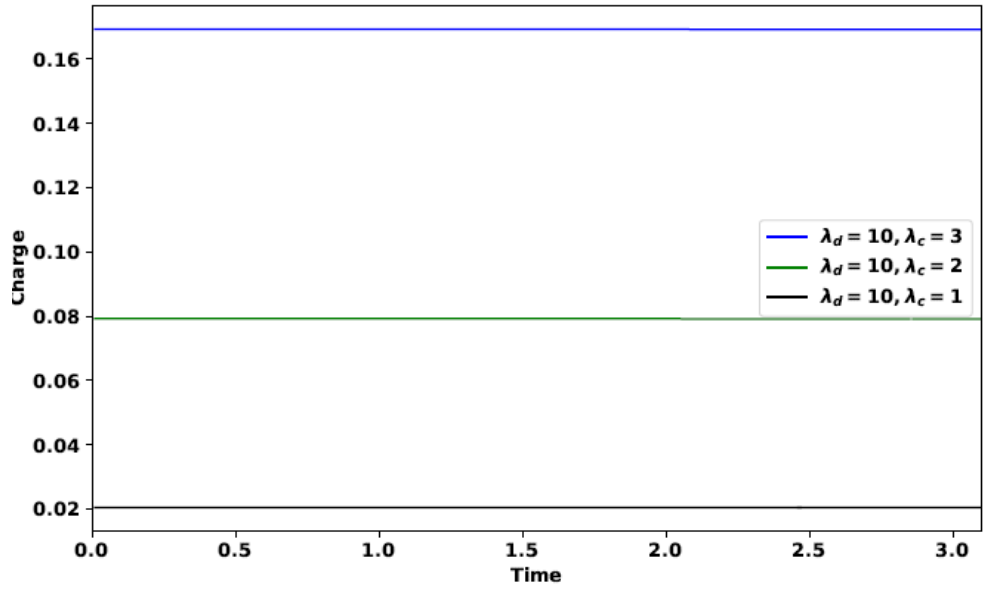

**Fig.4.** Average charge as a function of time in the Coulomb regime characterized by *P* = 100, 25, and 11.11 ($\overline{\lambda}_d$ = 10 and $\overline{\lambda}_c$ = 1, 2, 3). The magnitude of the average charge is constant up to fourth decimal place indicating that the system is in equilibrium.

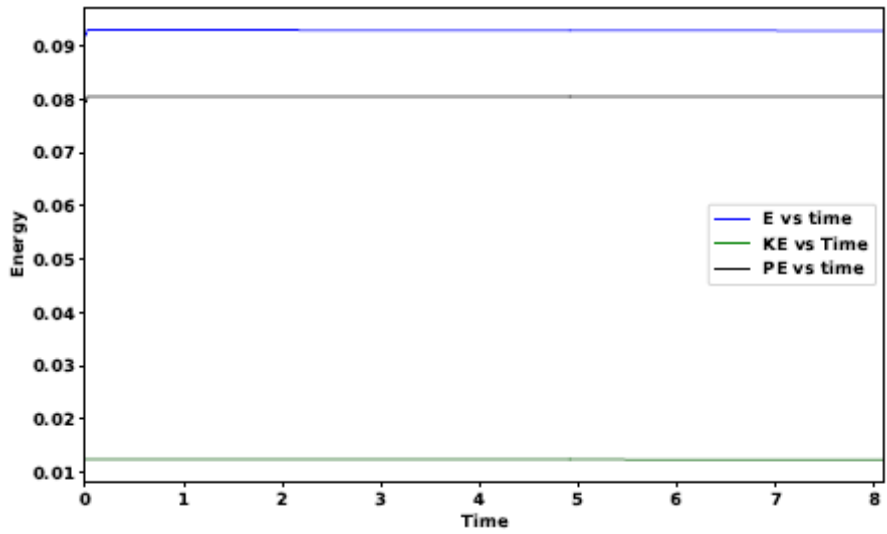

**Fig.5.** Energy conservation in the Coulomb regime for $P =11.11$ ($\bar{\lambda}_d = 10$ and $\bar{\lambda}_c = 3$). In this regime, the potential energy per particle is greater than the kinetic energy which is almost negligible. The total energy is conserved up to the fourth decimal place

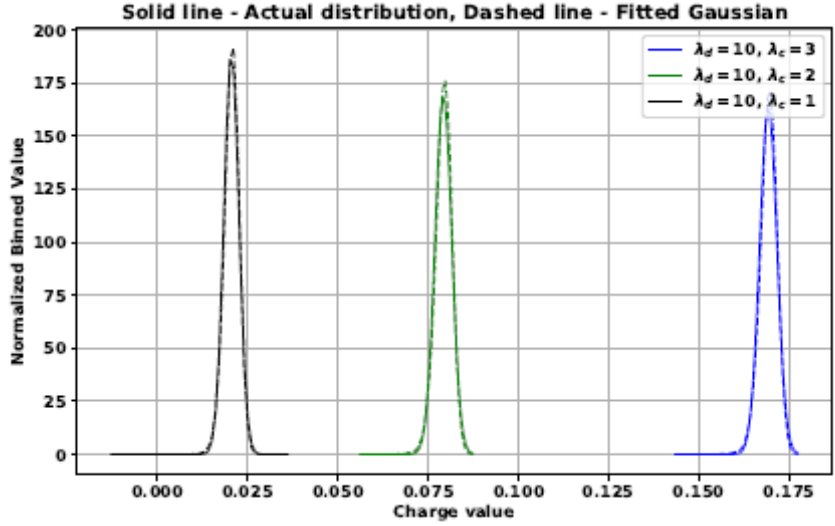

**Fig.6.** The charge distribution function $f_c$ in the Coulomb regime for P = 100, 25, 11.11 ($\overline{\lambda}_d = 1, \overline{\lambda}_c = 1,2,3$). The solid line is the distribution function from the simulation while the dashed line is the Gaussian fit. The fit shows that the $f_c$ is well approximated by a Gaussian distribution.

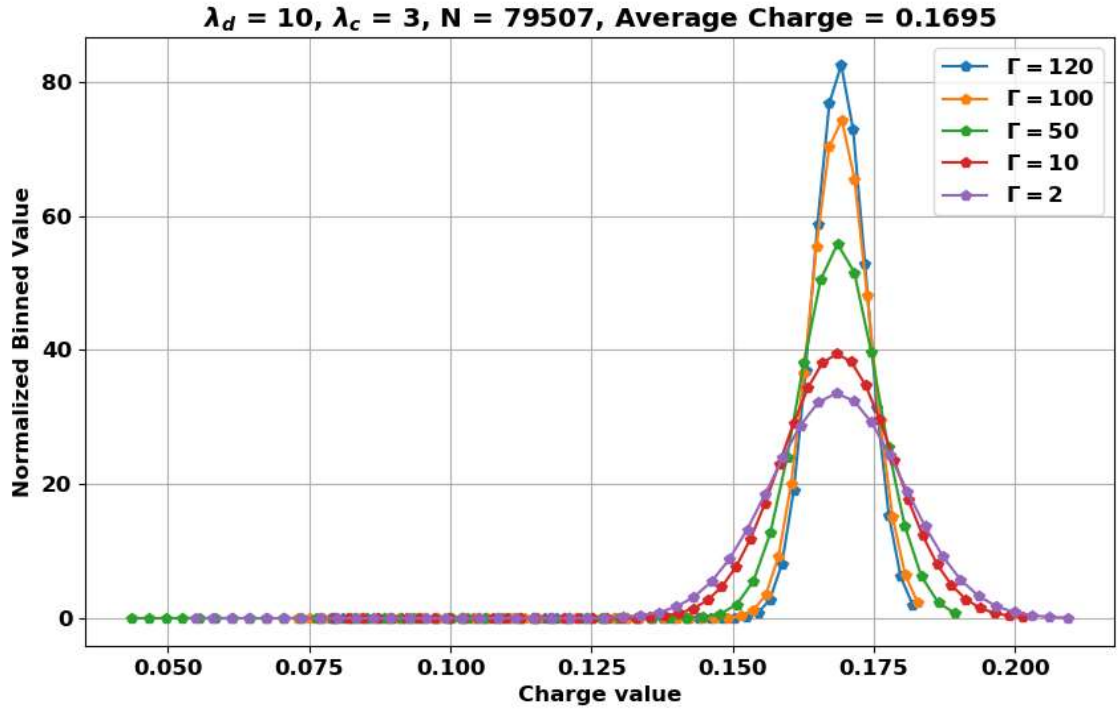

**Fig.7.** The charge distribution function $f_c$ in the Coulomb regime P = 11.11 as function of $\Gamma = 1/\overline{T}_d$ or the inverse dust temperature. The figure shows that with increasing $\overline{T}_d$ the width of $f_c$ increases around the fixed mean indicating the average dust charge is independent of its temperature.

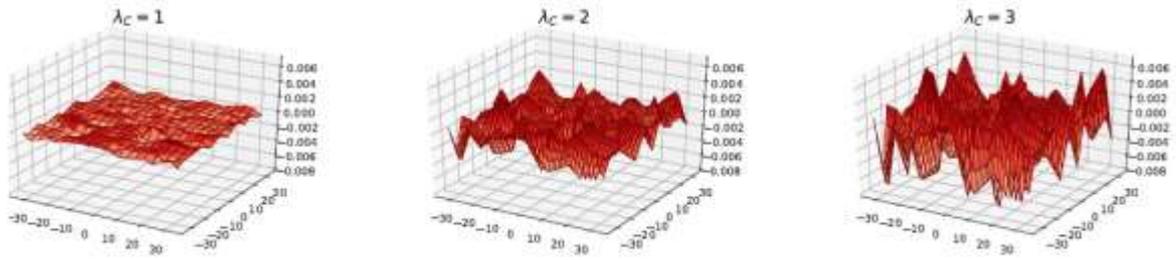

**Fig. 8.** 2D plots of $\rho_{dc}(x,z)$ in $y = 0$ plane for $\lambda_c = 1, 2, 3$ and $\lambda_d = 10$ and P = 100, 25 and 11.11. The plots clearly show that $\rho_{dc}$ becomes spatially uniform with increasing $P$. For $P = 100$, the charge density is nearly uniform.

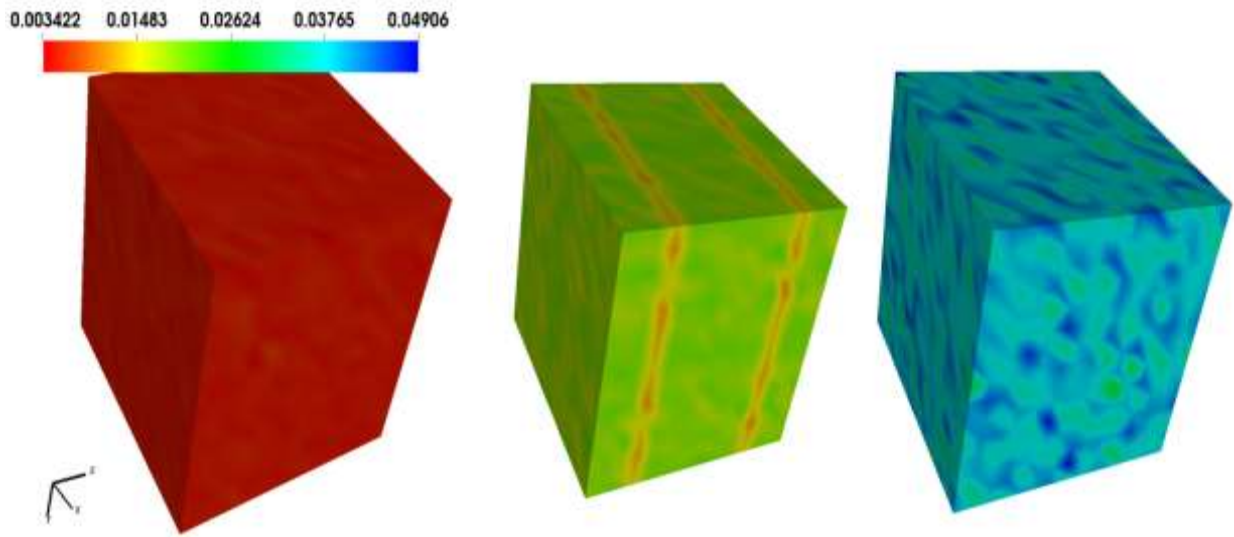

**Fig.9.** The figure shows color coded "charge density cubes" which are full 3D spatial plots of $\rho_{dc}(x, y, z)$ in the simulation box. The figure again shows the homogenization of dust charge density with increasing $P$. The first cube with $P = 100$ and normalised $\rho_{dc} = 1.84$ is homogenously colored indicating a uniform dust charge density. Hence this cube is well within the Coulomb regime hence may be called the " Coulomb plasma cube".